\documentclass[pdflatex,sn-mathphys]{sn-jnl}% Math and Physical Sciences Reference Style
\jyear{2021}%

\theoremstyle{thmstyleone}%
\theoremstyle{thmstyletwo}%

\theoremstyle{thmstylethree}%

\begin{document}

\title[Mock \textit{HUBS} observations]{Mock \textit{HUBS} observations of hot gas with IllustrisTNG}

%%=============================================================%%
%% Prefix	-> \pfx{Dr}
%% GivenName	-> \fnm{Joergen W.}
%% Particle	-> \spfx{van der} -> surname prefix
%% FamilyName	-> \sur{Ploeg}
%% Suffix	-> \sfx{IV}
%% NatureName	-> \tanm{Poet Laureate} -> Title after name
%% Degrees	-> \dgr{MSc, PhD}
%% \author*[1,2]{\pfx{Dr} \fnm{Joergen W.} \spfx{van der} \sur{Ploeg} \sfx{IV} \tanm{Poet Laureate} 
%%                 \dgr{MSc, PhD}}\email{iauthor@gmail.com}
%%=============================================================%%

\author[1]{\fnm{Yu-Ning}\sur{Zhang}}
%\email{zyn17@mails.tsinghua.edu.cn}

\author[1]{\fnm{Chengzhe} \sur{Li}}
% \equalcont{These authors contributed equally to this work.}

\author[1]{\fnm{Dandan} \sur{Xu}}
% \equalcont{These authors contributed equally to this work.}

\author*[1]{\fnm{Wei} \sur{Cui}}\email{cui@tsinghua.edu.cn}

\affil*[1]{\orgdiv{Department of Astronomy}, \orgname{Tsinghua University}, \orgaddress{\city{Beijing}, \postcode{100084}, \country{China}}}

% \affil[2]{\orgdiv{Department}, \orgname{Organization}, \orgaddress{\street{Street}, \city{City}, \postcode{10587}, \state{State}, \country{Country}}}

% \affil[3]{\orgdiv{Department}, \orgname{Organization}, \orgaddress{\street{Street}, \city{City}, \postcode{610101}, \state{State}, \country{Country}}}

%%==================================%%
%% sample for unstructured abstract %%
%%==================================%%

\abstract{The lack of adequate X-ray observing capability is seriously impeding the progress in understanding the hot phase of circumgalactic medium (CGM), which is predicted to extend to the virial radius of a galaxy or beyond, and thus in acquiring key boundary conditions for studying galaxy evolution. To this end, the Hot Universe Baryon Surveyor (\textit{HUBS}) is proposed. \textit{HUBS} is designed to probe hot CGM by detecting its emission or absorption lines with a non-dispersive X-ray spectrometer of high resolution and high throughput. The spectrometer consists of a $60\times60$ array of microcalorimeters, with each detector providing an energy resolution of $2~\mathrm{eV}$, and is placed in the focal plane of an X-ray telescope of $1\degr$ field-of-view. With such a design, the spectrometer is highly optimized for detecting X-ray-emitting hot gas in the CGM of local galaxies, as well as in intra-group medium (IGrM), intra-cluster medium (ICM), or intergalactic medium (IGM). To assess the scientific potential of \textit{HUBS}, in this work, we created mock observations of galaxies, groups, and clusters at different redshifts with the IllustrisTNG simulation. Focusing exclusively on emission studies, we took into account the effects of light cone, Galactic foreground emission, and background AGN contribution in the mock observations. From the observations, we made mock X-ray images and spectra, analyzed them to derive the properties of the emitting gas in each case, and compared the results with the input parameters from the simulation. The results show that \textit{HUBS} is well suited for studying hot CGM at low redshifts. The redshift range is significantly extended for measuring IGrM and ICM. The sensitivity limits are also presented for detecting extended emission of low surface brightness. }


%%\pacs[JEL Classification]{D8, H51}

%%\pacs[MSC Classification]{35A01, 65L10, 65L12, 65L20, 65L70}

\maketitle

\section{Introduction}
\label{sec:intro}

Our understanding of galaxies has evolved significantly over the past decade or so. Stimulated by multi-wavelength observations and theoretical studies, we began to view a disc galaxy in terms of a complex ecosystem \citep[see][for a recent review]{2017ARA&A..55..389T}: gas in the disc may be heated and expelled into the halo, likely through the action of stellar and/or supermassive black hole feedback, and may subsequently cool and condense in the halo, and eventually, together with newly accreted gas from the larger-scale inter-galactic medium (IGM), fall back onto the disc, providing cold gas for fueling the formation of the next-generation of stars. Though varying in details, such a cycling of baryonic matter between the disc and halo has become a fairly generic feature in theoretical models, but is still seeking observational confirmation. The studies of circum-galactic medium (CGM) have gained significant momentum in recent years\footnote{\url{https://www.kitp.ucsb.edu/activities/halo21}}, but there are still many unresolved issues remaining, which impede the progress in our understanding of galaxy evolution. 

Hydrodynamical simulations show that the CGM is inherently multi-phase, which has also found significant observational support. But, exactly how different phases of the CGM are mixed and distributed is still a matter of much debate, due to the lack of relevant observations, especially X-ray observations that probe the hot phase, which is predicted to be very extended spatially. The observations of edge-on galaxies have established the presence of X-ray-emitting gas in the halo \cite[e.g.][]{2001ApJ...555L..99W, 2004ApJS..151..193S, 2006A&A...448...43T, 2013MNRAS.435.3071L, 2017ApJS..233...20L}, 
but only on scales of the order $10~\mathrm{kpc}$, which are quite small when compared with virial radii of typically $200\text{--}300~\mathrm{kpc}$. If such hot gas extends up to the virial radius of galaxies or beyond, as some models suggest, it could help solve the problem of baryon deficit in galaxies like the Milky Way \cite[e.g.][]{2011ApJ...737...22A, 2012ApJ...755..107D, 2017ApJ...850...98B}, and also provide the boundary conditions necessary for understanding the physics of feedback and physical processes in the CGM.   

Directly observing X-ray emission from the CGM is challenging, especially in the outer part, as the density (and thus the emission measure) is likely quite low. On the other hand, for an optically-thin, collisionally-ionized plasma, the emission spectrum is expected to be dominated by lines, so high-resolution X-ray spectroscopy could be an effective way to extend the reach of CGM studies. In this context, the Hot Universe Baryon Surveyor (\textit{HUBS})\footnote{\url{http://hubs.phys.tsinghua.edu.cn/en/index.html}} mission was proposed (\cite{2020JLTP..199..502C}; see updates in \cite{2020SPIE11444E..2SC}). \textit{HUBS} aims at filling an void in the detection of X-ray emission from the hot phase of the CGM by employing a non-dispersive X-ray spectrometer of high spectral resolution and high throughput. In this work, we assess the scientific capabilities of \textit{HUBS} related to the studies of CGM, as well as of hot gas in galaxy groups or clusters, with mock observations.

The paper is structured as follows: In Section \ref{sec:HUBS}, we provide a brief description of the \textit{HUBS} mission, followed by a detailed description of the methodology for creating mock X-ray observations in Section \ref{sec:Methods}. The results from analyzing the mock observations of galaxies, groups, and clusters are shown in Section \ref{sec:analysis}. We discuss the results and conclude in Section \ref{sec:summary}.

\section{The \textit{HUBS} payload}
\label{sec:HUBS}

The design of the \textit{HUBS} payload is highly optimized for the detection of extended X-ray emission of low surface brightness \cite{2020SPIE11444E..2SC}. It puts a strong emphasis on the softest X-ray band (below about $1~\mathrm{keV}$), in which the emission lines of the hot CGM are expected to be mainly located, but extends the spectral coverage to about $2~\mathrm{keV}$. The X-ray focusing optics employed features a large field-of-view (FoV) to maximize the throughput of detecting diffuse X-ray emission, which is often quantified by the product of the effective area and FoV (often referred to as $\it grasp$). At the focal plane of the optics lies an array of microcalorimeters, which are based on the technology of superconducting transition-edge sensors, with their quantum efficiency (QE) and energy resolution optimized for operating at below $1~\mathrm{keV}$. A sub-array of smaller detectors are placed at the centre of the main detector array, providing even higher energy resolution. Such a hybrid design of the detector array is adopted to enhance absorption-line observations of bright background (point) sources. The key parameters of the \textit{HUBS} payload are summarized in \cite[Table~1]{2020SPIE11444E..2SC}. 

The throughput of the system is not only determined by the collecting area of X-ray optics, but also by the transmission coefficients of the blocking filters and by the QE of the detectors. The blocking filters are placed in front of the detector array to reject photons at long wavelengths, primarily infrared and optical photons, and thus to reduce shot noise, which would degrade the energy resolution of the detectors. For this work, the collecting area adopted is based on a preliminary design of the optics, as provided by the optics development team (Z.-S. Wang, private communication). To take into account optical vignetting, we averaged the data over a range of off-axis angles up to $30\arcmin$ at each X-ray energy, to derive the FoV-averaged collecting area curve. We note that efforts are being made by the optics development team to further improve the performance of the optics below $1~\mathrm{keV}$. 

As for the response of the filters, we adopted the measured transmission curve of the blocking filters used by the \textit{XQC} sounding rocket experiment (D. McCammon, private communication; also see \cite{2002ApJ...576..188M}). Deviating from the five-filter system of \textit{XQC}, we plan to use six filters in \textit{HUBS}, with the outermost one located at room temperatures for minimizing contamination of the filter/detector system, so we scaled the \textit{XQC} filter transmission curve accordingly. Here, we assumed a detector QE of unity. 

\begin{figure}[h]
    \centering
	\includegraphics[width=0.5\textwidth]{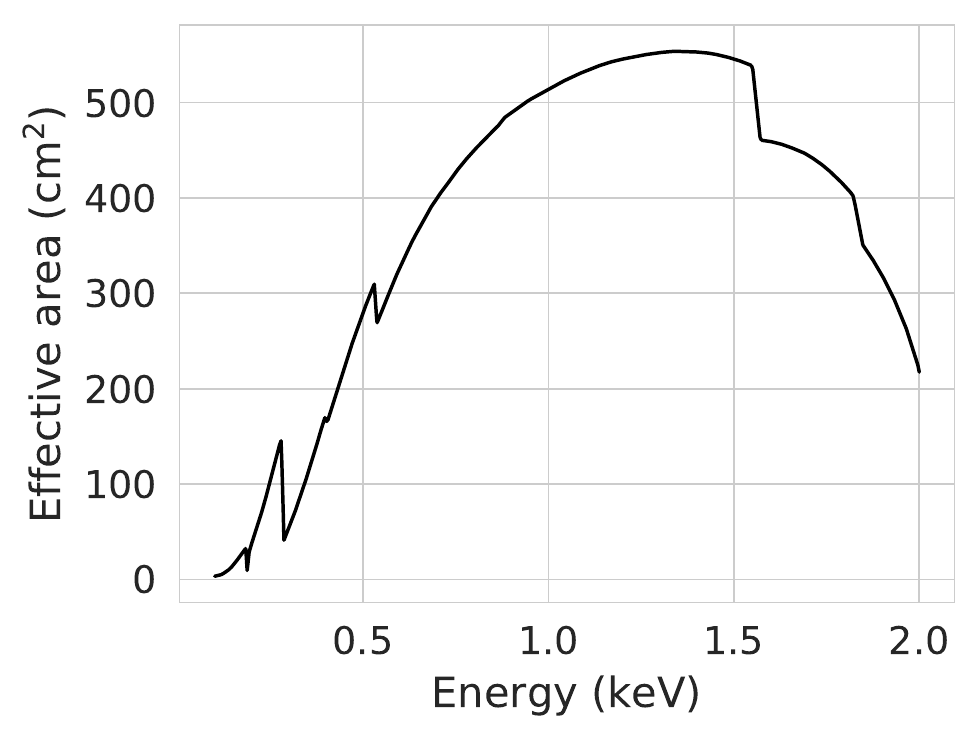}
    \caption{Preliminary effective area of \textit{HUBS}, as a function of X-ray energy. The effects of optical vignetting have been considered (see text). }
    \label{fig:arf}
\end{figure}

For creating mock observations, we ignored detector background caused by energetic charged particles in the space environment. This is justified by the fact that \textit{HUBS} is designed to operate in a low-earth orbit, where the particle background is relatively low, especially in the soft X-ray band that \textit{HUBS} covers, and is expected to be negligible compared with the X-ray background of cosmic origin (see Sec.~\ref{sec:bgs}). Multiplying the collecting area of the optical system and the transmission of the filters, we computed the overall effective area of the system, as shown in Fig.~\ref{fig:arf}. Then, we generated the corresponding ancillary response file (ARF), as well as a response matrix file (RMF), for subsequent spectral analyses. For the RMF, we divided the \textit{HUBS} passing band ($0.1\text{--}2~\mathrm{keV}$) into $9500$ pulse height channels (with each $0.2~\mathrm{eV}$ in width). The line response of the detector was taken to be of Gaussian shape with its full-width-at-half-maximum (FWHM) set at $2~\mathrm{eV}$. Note that we have neglected the difference of the central sub-array of detectors in this work, as we focus exclusively on emission lines.

\section{Mock observation}
\label{sec:Methods}

The procedure for producing mock X-ray observations is similar to that of past studies (e.g., \cite{2003PASJ...55..879Y, 2005ApJ...623..612F, 2012MNRAS.424.1012R, 2020ApJ...893L..24O}). 

\subsection{Cosmological simulation}
\label{sec:tng}
We utilized data from the IllustrisTNG simulation \footnote{\url{http://www.tng-project.org/data/}} ( \cite{Marinacci_et_al.(2018),Naiman_et_al.(2018),Nelson_et_al.(2018),Nelson_et_al.(2019a),Pillepich_et_al.(2018b),Springel_et_al.(2018)}), 
which was carried out using an advanced moving-mesh code \textsc{arepo} \cite{Springel(2010)}. 
For making mock observations, we made use of the full-physics run (TNG100), which has a cubic box volume of $(110.7\ \mathrm{Mpc})^3$, a mass resolution of $1.4\times10^6\ {\rm M_{\sun}}$ and $7.5\times10^6\ {\rm M_{\sun}}$ for the baryonic and dark matter, respectively, and a gravitational softening length of $0.5\ h^{-1}\,\mathrm{kpc}$. Galaxies hosted by their dark matter haloes were identified with the {\sc subfind} algorithm \cite{Springel_et_al.(2001),Dolag_et_al.(2009)}. The simulation adopted the Planck cosmology parameters under the assumption of a flat universe geometry \cite{Planck_Collaboration(2016)}: a matter density of $\Omega_{\rm m} = 0.3089$ (with a baryonic density of $\Omega_{\rm b} = 0.0486$), a cosmological constant of $\Omega_{\Lambda} = 0.6911$, and a Hubble constant $h = \left.H_0\right/\left(100\ {\rm km\,s}^{-1}\,{\rm Mpc^{-1}}\right) = 0.6774$. This cosmology was also assumed throughout this work.
 
\subsection{Target selection}
\label{sec:source}

Because the spectral coverage of \textit{HUBS} is limited to the soft X-ray band, it primarily targets on the hot phase of the CGM, as well as hot gas in the intra-group medium (IGrM) or in the outskirts of the intra-cluster medium (ICM), as well as in the filamentary structures. For this work, we selected Friends-of-Friends \cite{1985ApJ...292..371D} haloes in the TNG100 simulation at three mass scales, including galaxies ($10^{12}~\mathrm{M_{\sun}} < M<10^{13}~\mathrm{M_{\sun}}$), galaxy groups ($10^{13}~\mathrm{M_{\sun}} < M<10^{14}~\mathrm{M_{\sun}}$), and galaxy clusters ($M > 10^{14}~\mathrm{M_{\sun}}$). To characterize the spatial extent of the dark matter halo of a target source, we define its virial radius as the radius of a sphere in which the mean density is equal to $200$ times the critical density of the universe ($r_{200}$). To separate the central region of the halo from its outskirt, we have further defined a fiducial radius $r_{500}$ in a similar manner. 

\begin{figure}[h]
	\centering
	\includegraphics[width=\textwidth]{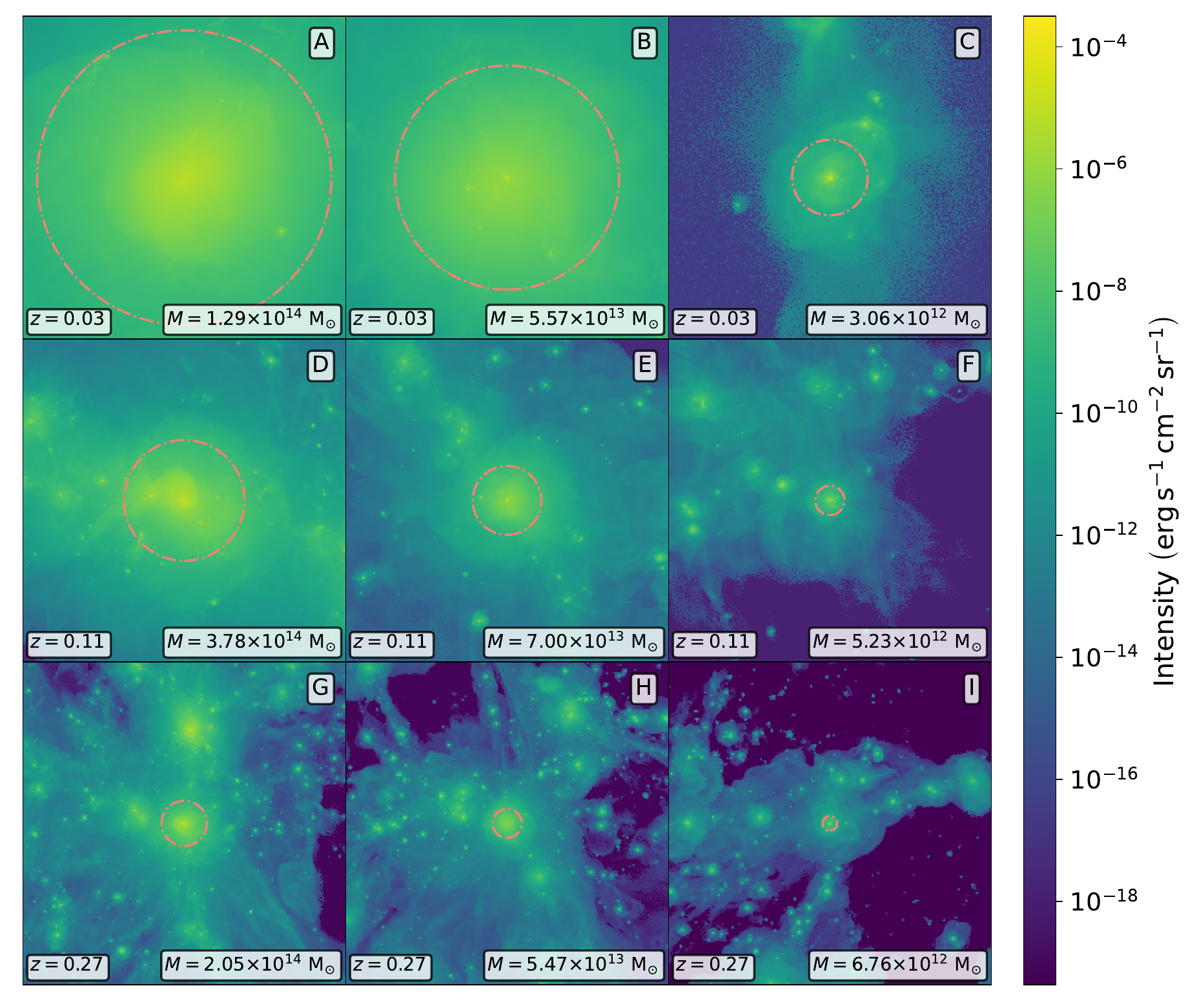}
    \caption{X-ray emissivity maps of the selected targets. Each image is $256 \times 256$ in dimension, covering a sky region of $1\degr \times 1\degr$. The dash-dot circle is of radius $r_{200}$. Note that zero-valued pixels were set to the minimum value of a map. }
    \label{fig:intensity}
\end{figure}

After going through the snapshots that cover the redshift range of interest, we found many candidate haloes corresponding to the three mass scales for galaxies, galaxy groups, and galaxy clusters. For simplicity, we selected relatively isolated dark matter haloes at each mass scale for three redshifts ($z=0.03$, $0.11$, and $0.27$). 
The properties of the targets are summarized in Table~\ref{tab:sources}. The corresponding X-ray emissivity maps of the targets were computed and are shown in Fig.~\ref{fig:intensity}. Note that the redshift of a target is set to be that of the snapshot in which it is located.

As a caveat, we note that the observed metallicity in rich clusters stays nearly constant around $0.2-0.4 Z_{\sun}$ outside $\sim 0.15 r_{200}$ (e.g., \cite{Simionescu2011, Gastaldello2021}). For comparison, we made metallicity profiles for $16$ TNG100 cluster samples at $z=0$ and averaged them to arrive at a profile typical of simulated clusters. The average profile shows that the metalicity decreases towards the outskirt and reaches about $0.2 Z_{\sun}$ at $r_{200}$, which is lower than the observed values. This suggests that real systems might be brighter in metal emission lines than the simulated ones.

\begin{table}
	\centering
	\caption{Properties of the selected targets$^\dag$}
	\label{tab:sources}
	\begin{tabular}{cccccc} % four columns, alignment for each
		\toprule%
		Target & Redshift & Mass & $M_{200}$ & $r_{200}$ & Type \\
        ID & & $\left(\mathrm{M_{\sun}}\right)$ & $\left(\mathrm{M_{\sun}}\right)$ & $\left(\mathrm{Mpc}\right)$ & \\
		\midrule
		A & $0.03$ & $1.29\times10^{14}$ & $1.16\times10^{14}$ & $1.02$ & cluster \\
        B & $0.03$ & $5.57\times10^{13}$ & $5.12\times10^{13}$ & $0.77$ & group \\
        C & $0.03$ & $3.06\times10^{12}$ & $1.98\times10^{12}$ & $0.26$ & galaxy \\ 
		D & $0.11$ & $3.78\times10^{14}$ & $3.21\times10^{14}$ & $1.39$ & cluster \\
        E & $0.11$ & $7.00\times10^{13}$ & $5.95\times10^{13}$ & $0.79$ & group \\
        F & $0.11$ & $5.23\times10^{12}$ & $4.60\times10^{12}$ & $0.34$ & galaxy \\
        G & $0.27$ & $2.05\times10^{14}$ & $1.77\times10^{14}$ & $1.08$ & cluster \\
        H & $0.27$ & $5.47\times10^{13}$ & $4.66\times10^{13}$ & $0.69$ & group \\
        I & $0.27$ & $6.76\times10^{12}$ & $6.04\times10^{12}$ & $0.35$ & galaxy \\        
		\botrule
		\multicolumn{6}{l}{$^\dag$ An exposure time of $1~\mathrm{Ms}$ is adopted for making mock observations.} \\
	\end{tabular}
\end{table}

\subsection{Light cone}
\label{sec:lc}

In reality, along the line of sight towards a target, other sources (e.g., star-forming galaxies or clusters) could also produce X-ray emission lines in the energy range of interest. To mimic actual observations, light cones need to be generated and used when creating mock observations. Along a line of sight, only photons from within the corresponding light cone can reach the detector.

In order to generate light cones from the simulated data without repeated structures, we used the method of \cite{2007MNRAS.376....2K}. The key to this method is to use a pair of integers $(n,m)$ that have no common denominator. 
The details of the method can be found in \cite[Section 2.4.1]{2007MNRAS.376....2K}.
Briefly, the integer pair sets the direction vector of a line of sight as $\mathbf{u_3}=\left.\left(n,m,mn\right)\right/\left(n^2+m^2+m^2n^2\right)^{\left.1\right/2}$.
The light cone along the line of sight would be ensured of no repeating structures from the origin to the positions at coordinates $\left(\left(n \pm \left.0.5\right/m\right)L,\left(m \pm \left.0.5\right/n\right)L,nmL\right)$ in the data cube.
The angular area covered by the light cone is about $\left.1\right/ \left(m^2n\right) \times \left.1\right/ \left(n^2m\right)~\mathrm{\left(rad^2\right)}$.
In other words, for a chosen pair of integers, the redshift range and angular area of the light cone are uniquely determined. 

Considering the size of TNG100 data cubes and the design parameters of the \textit{HUBS} payload, we chose $(n,m)=(4,3)$ to generate a light cone, corresponding to an angular area of $\sim1.59\degr \times 1.19\degr$ (which is slightly larger than the FoV of \textit{HUBS}) and a redshift range of roughly $0\text{--}0.356$. For making mock light-cone observations, we stacked the TNG100 snapshots at $z=0.01$, $0.03$, $0.06$, $0.11$, $0.18$, $0.27$, and $0.35$.
Through swapping the integers $(n,m)$, changing the origin (e.g. $(0,0,0) \mapsto (L,0,0)$, where $L$ is the side of the box), and permutating coordinate axes (e.g. $x \mapsto y$, $y \mapsto z$ and $z \mapsto x$), $48$ light cones were generated. For this work, we took the one that contains no extremely bright structures, compared with the selected targets (as shown in Table~\ref{tab:sources}). 

We used the same light cone to generate mock observations of all targets. Briefly, to produce a mock observation of a target, the light cone was divided into redshift slices. Each slice was treated as an individual observation at one redshift, and all slices are integrated to produce the overall X-ray emission along the light cone. 
%this adds a line-of-sight contribution to the observation of the target. 
Then, we put the target in the light cone, calculated its X-ray emission, and added it, along with the foreground and background emission (see Sec.~\ref{sec:bgs}), to the light-cone contribution, producing the final mock observation.

\subsection{Foreground and background}
\label{sec:bgs}

Hot gas also exists in our own Galaxy. It is often characterized in terms of the so-called ``Local Hot Bubble'' and Galactic halo. In addition, the ions in the solar wind can exchange electric charge with neutral atoms in the interstellar medium or in the Earth exosphere, to produce X-ray emission whose spectrum is also dominated by emission lines. Together, they form a foreground for observing an extragalactic source.

For this work, the X-ray foreground was generated with the \texttt{SOXS} (v3.0.2) package\footnote{\url{https://hea-www.cfa.harvard.edu/soxs/index.html}}. The X-ray spectrum was calculated with the \textsc{apec} code (v3.0.9; \cite{2012ApJ...756..128F}), assuming the hot gas is in collisional ionization equilibrium (which is generally the case for the sources of interest here). We used two \textsc{apec} components to approximate the foreground emission\footnote{\url{https://hea-www.cfa.harvard.edu/soxs/users_guide/background.html}}. 

The X-ray background mainly originates in unresolved active galactic nuclei (AGN) and, to a lesser extent, normal galaxies. The latter are already included in the light-cone component. To compute the AGN contribution, we used
the $\log N - \log S$ relation \cite{2012ApJ...752...46L} to create a population of background AGN. 
Each source was assumed to have an X-ray spectrum of power-law shape, with photon indices drawn from the distribution of measured values \cite[see Fig. 13a of][]{2006ApJ...645...95H}.
The X-ray spectrum of the AGN background was generated also with the \texttt{SOXS} package. 

\subsection{Summary of procedures}
\label{sec:tools}

In summary, the steps for generating mock observations are as follows:

\begin{itemize}
  \item Selecting a halo of mass appropriate for the target of interest from the TNG100 data, for a given redshift, and placing it at the centre of a sky field of roughly $1\degr \times 1\degr$ in size.
  \item Computing the X-ray emission using the \texttt{pyXSIM} package\footnote{\url{http://hea-www.cfa.harvard.edu/~jzuhone/pyxsim/}}, based on the density, temperature, metallicity of each gas element in the target, under the assumption that the X-ray spectrum is described by the \textsc{apec} model. The X-ray photons was then redshifted appropriately, according to the redshift of the target. In this step, we assumed an exposure time of $1~\mathrm{Ms}$ and a collecting area of $A_\text{coll} = 1000~\mathrm{cm^2}$ for generating X-ray photons.    
  \item Projecting the X-ray photons along the line of sight to the target. Based on the peculiar velocity of each particle, the wavelength of X-ray photons is Doppler-shifted (in addition to the cosmological redshift). To relate to observations, foreground absorption was implemented with the \verb"tbabs" model \cite{2000ApJ...542..914W}, assuming an interstellar hydrogen column density of $N_{\rm{H}} = 4 \times 10^{20}~ \mathrm{cm^{-2}}$.
  \item Making a light cone to account for X-ray emission from sources other than the target along the line of sight. We computed X-ray emission from all objects that lie inside the light cone.
  \item Generating X-ray foreground and background for the mock observation.
  \item Convolving the overall X-ray emission (from the target, light cone, AGN background, and Galactic foreground) with the instrument responses (RMF and ARF) using the instrument simulator module in the \texttt{SOXS} package, to generate the detected events. 
  \item Binning the events in spatial and spectral coordinates, respectively, to construct mock X-ray images and spectra.

\end{itemize}

\section{Analyses and Results}
\label{sec:analysis}

\subsection{Imaging analyses}
\label{sec:image}

\begin{figure}[ht]
    \centering
	\includegraphics[width=\textwidth]{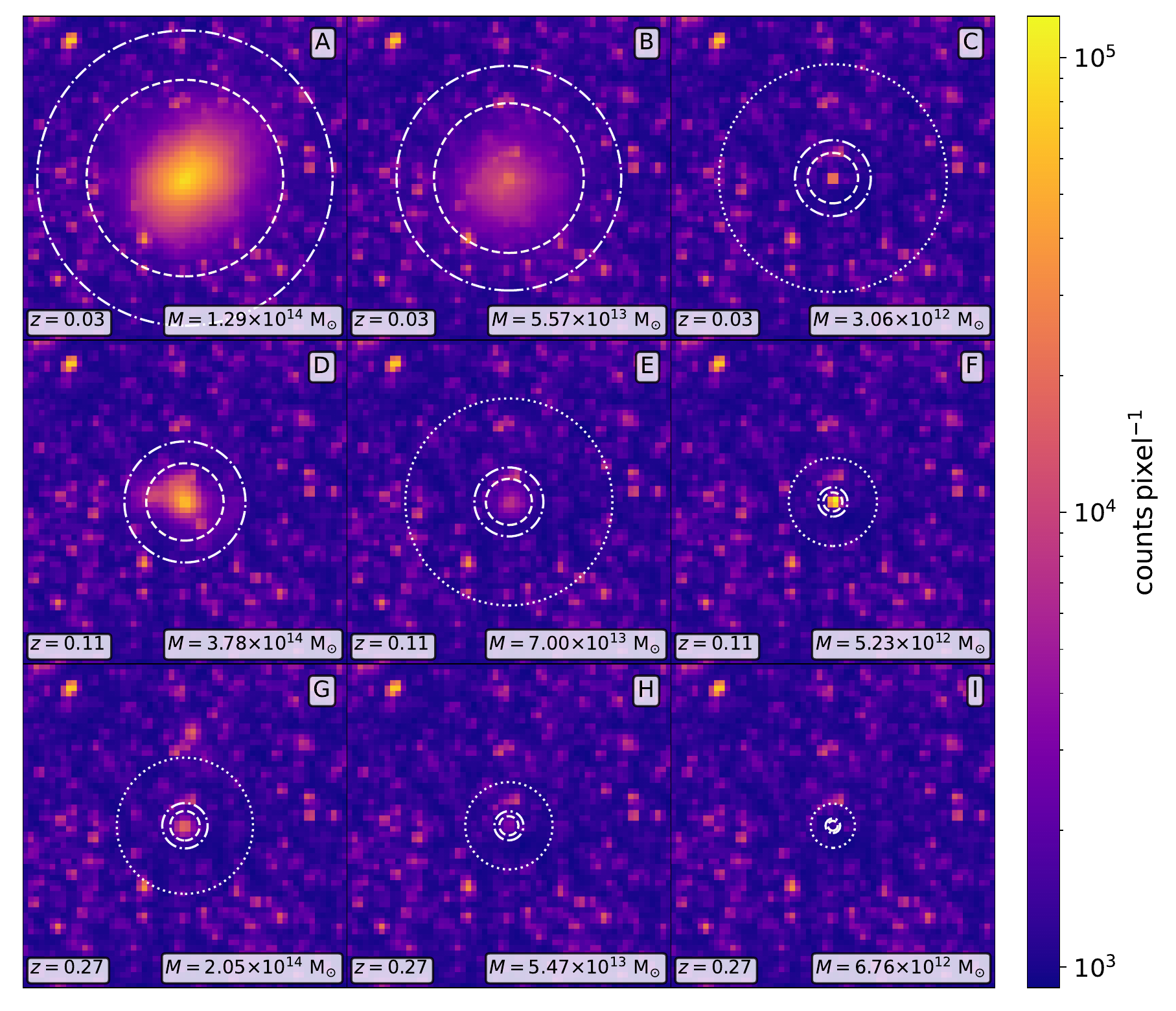}
    \caption{Mock 0.1-2 keV X-ray images of the selected targets, with an assumed exposure time of $1~\mathrm{Ms}$. The dashed, dot-dashed and dotted circles are of radii $r_{500}$, $r_{200}$, and $3r_{200}$, respectively. Each image is $60 \times 60$ in dimension  (with $1\arcmin$/pixel), covering a sky region of $1\degr \times 1\degr$. }
    \label{fig:images}
\end{figure}

Fig.~\ref{fig:images} shows the images derived from the mock observations. Much of the large-scale filamentary structures (as seen in the emission maps in Fig.~\ref{fig:intensity}) is no longer discernible in the images, but X-ray emissions from CGM, IGrM or ICM of the target source are detected.

To emulate data analyses with real observations, we used the \textit{DAOStarFinder} tool in the \textsc{photutils} package\footnote{\url{https://photutils.readthedocs.io/en/stable/}} to detect point-like sources (i.e., the background AGN) in the images. For that, a 2-d Gaussian kernel of radius $1.5$ pixels (FWHM) was employed, with the detection threshold set at $3\sigma$ (above the local background). The detected point-like sources were removed by masking out a circular area of diameter $3$ pixels around each source. The blank pixels were then filled with values interpolated from neighboring pixels using the \textit{interpolate\_replace\_nans} script (in \textsc{astropy}\footnote{\url{http://www.astropy.org}}). 

\begin{figure}[ht]
    \centering
	\includegraphics[width=0.5\textwidth]{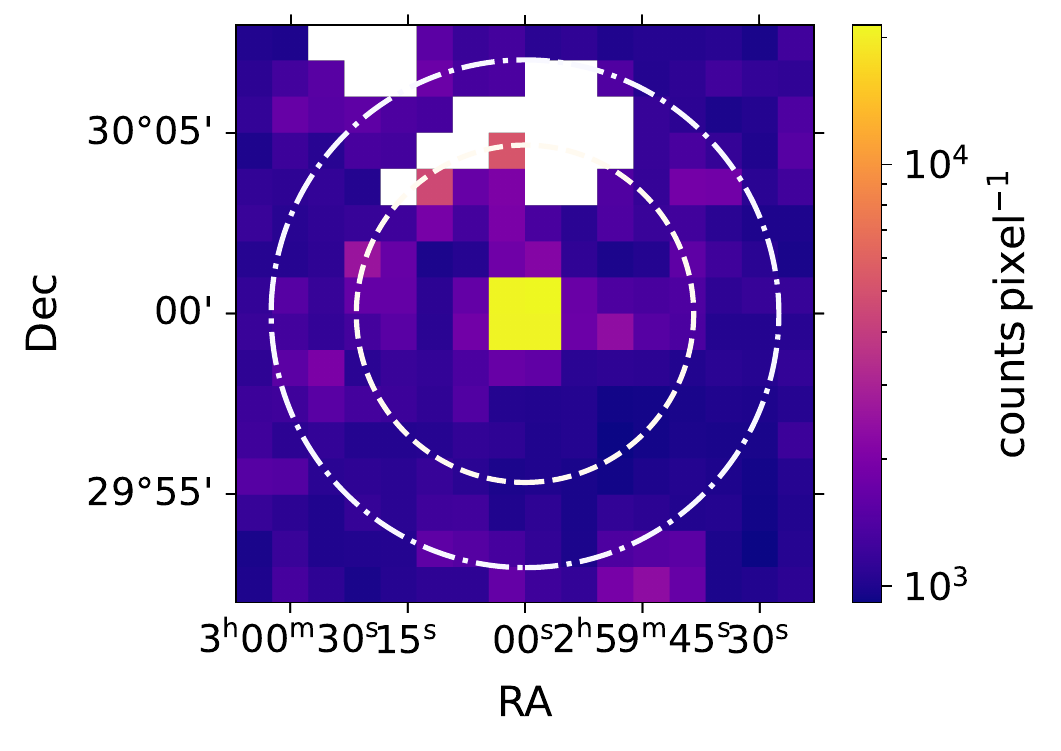}
    \caption{Zoomed-in image of Target C. The blank pixels are associated with the removed sources. The dashed and dot-dashed circles show regions of radius $r_{500}$ and $r_{200}$, respectively. }
    \label{fig:zoom}
\end{figure}

Then, We proceeded to detect and remove contamination from extended sources (e.g. star-forming galaxies) in the light cone, with the \textit{detect\_sources} tool in the \textsc{photutils} package, with the detection threshold set at $3\sigma$ (and the npixels parameter set to 5). As an example, Fig.~\ref{fig:zoom} shows a cleaned image around a target, where the blank pixels are associated with extended contaminating sources. 

\begin{figure}[h]
    \centering
	\includegraphics[width=\textwidth]{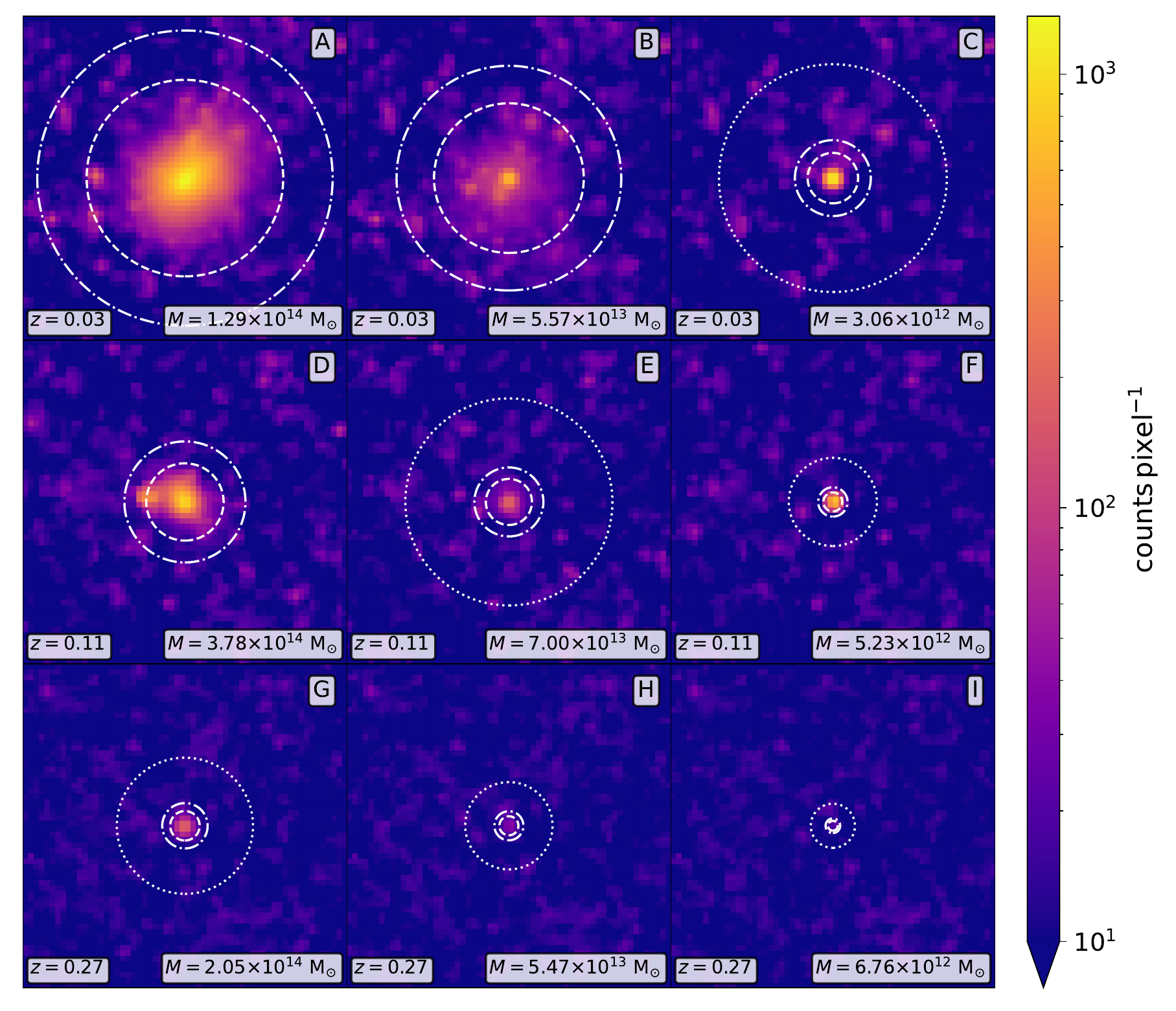}
    \caption{As for Fig.~\ref{fig:images}, but in a narrow band around the (redshifted) \ion{O}{\romannumeral7} emission line. The width of the band was chosen to be about 18 eV, to ensure the inclusion of the line triplets. The point-like and extended sources detected in the full-band images were removed and the images smoothed. }
    \label{fig:images_ovii}
\end{figure}

\begin{figure}[h]
    \centering
	\includegraphics[width=\textwidth]{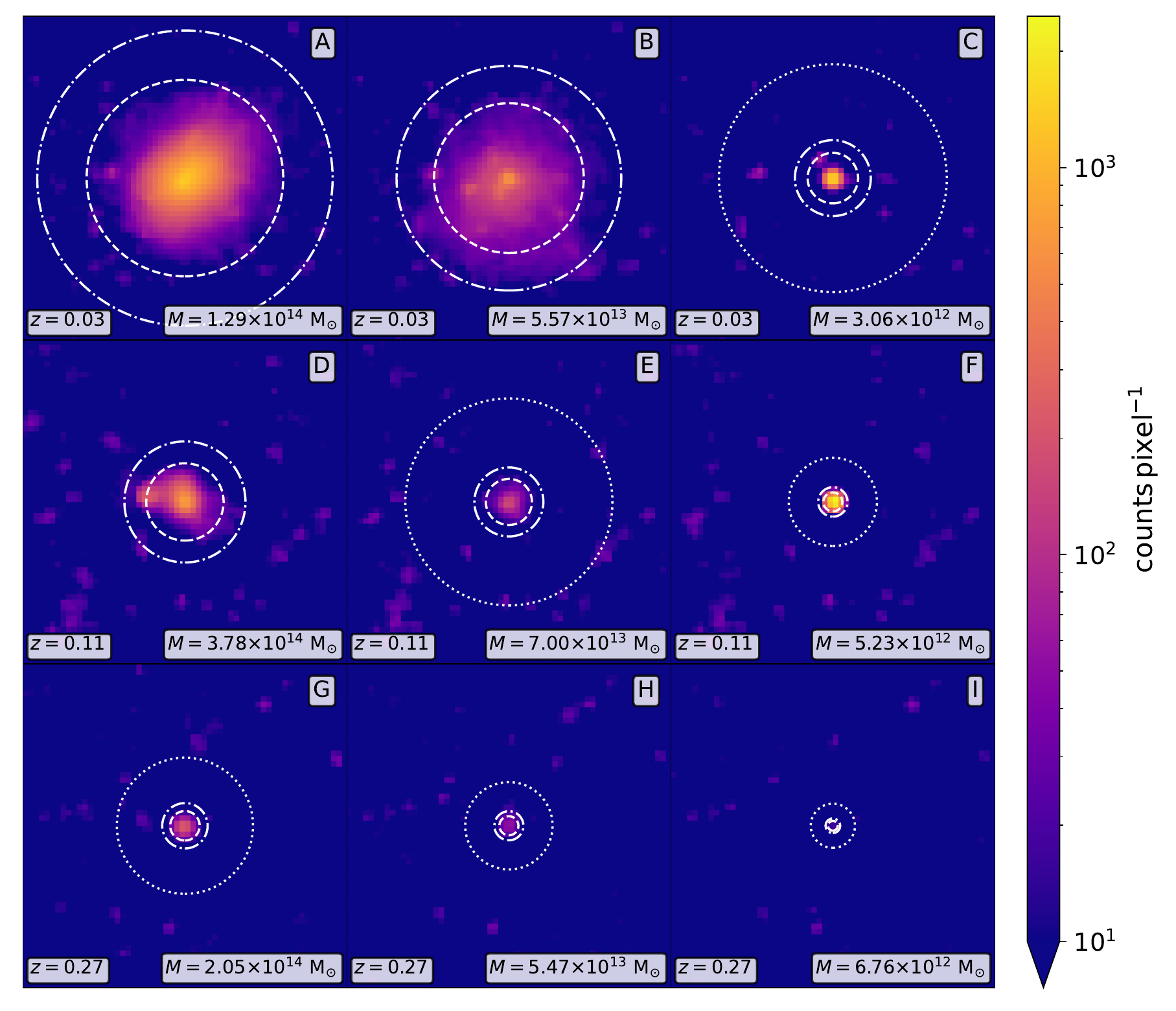}
    \caption{As for Fig.~\ref{fig:images_ovii}, but in a narrow band around the (redshifted) \ion{O}{\romannumeral8} emission line. }
    \label{fig:images_oviii}
\end{figure}

To improve the signal-to-noise ratio of spatial analyses, we also made images using the same approach as described above but in a narrow band around the bright (red-shifted) \ion{O}{\romannumeral7} and \ion{O}{\romannumeral8} emission lines, respectively. 
The results are given in Figs.~\ref{fig:images_ovii} and \ref{fig:images_oviii}. As can be seen, the extended X-ray emission from CGM stands out more significantly in the narrow-band images (in particular, around \ion{O}{\romannumeral8}); so does that of IGrM (for groups) and ICM (for clusters). 

\begin{figure}[ht]
    \centering
	\includegraphics[width=1.\textwidth]{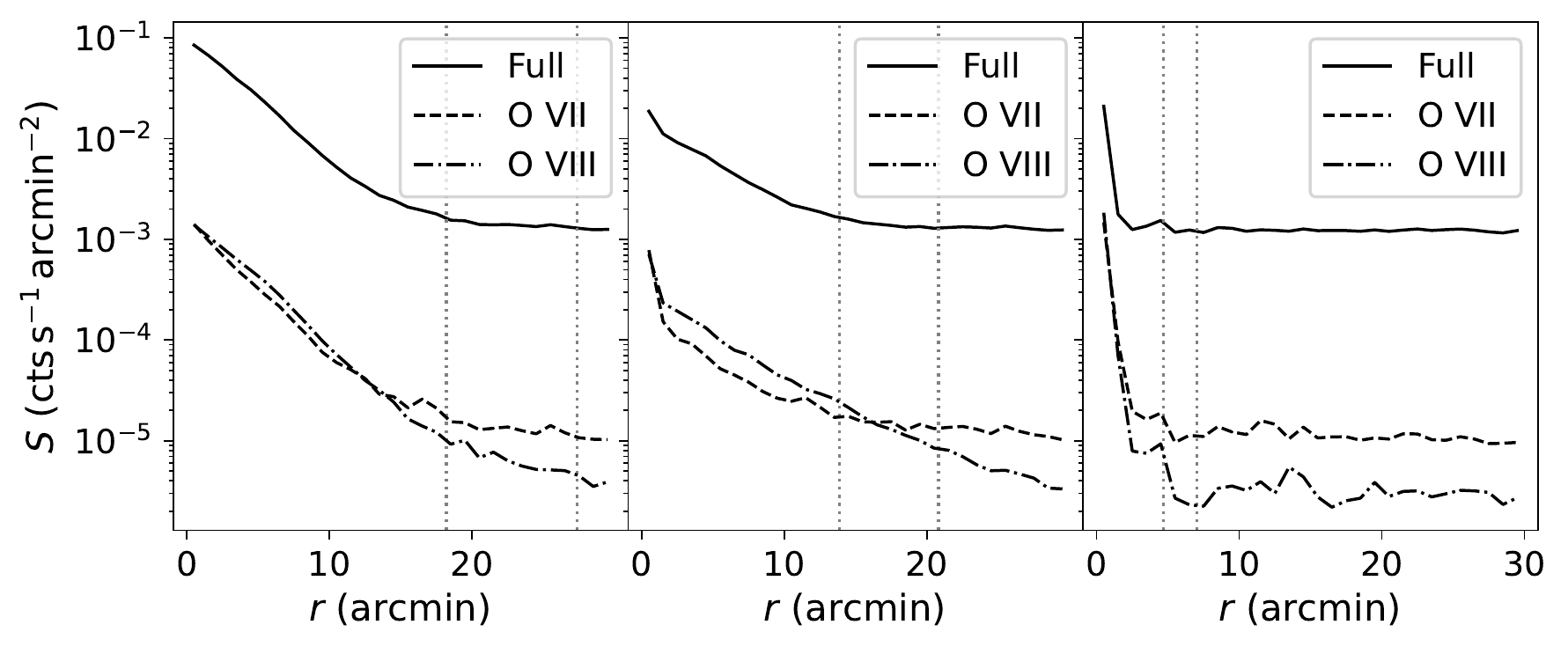}
    \caption{The radial profiles of Target A ({\it left}), Target B ({\it middle}), and Target C ({\it right}), respectively. As indicated, the profiles were derived from the full-band and narrow-band images, and are shown separately. The dotted lines indicate $r_{500}$ and $r_{200}$.}
    \label{fig:profile}
\end{figure}

To be more quantitative, we made radial profiles from the images. Examples are shown in Fig.~\ref{fig:profile} for the cluster, group, and galaxy, respectively, at redshift $z=0.03$. For both the cluster and group, while the narrow-band profile generally follows the full-band profile around the \ion{O}{\romannumeral7} line, roughly to $r_{500}$, it is significantly steeper than the full-band profile around the \ion{O}{\romannumeral8} line and extends beyond $r_{200}$. The increase in the ratio of \ion{O}{\romannumeral7} and \ion{O}{\romannumeral7} surface brightness would be consistent with a cooler outskirt in these systems. The galaxy is a bit difficult to resolve even at such a low redshift, which indicates that the primary targets for CGM studies with \textit{HUBS} are nearby galaxies. In this case, it can still be seen that the narrow-band profiles are more extended than the full-band profile.

The results indicate that \textit{HUBS} would be quite effective in detecting emission from hot CGM, through direct imaging, but may be limited to nearby (or low-redshift) galaxies.
IGrM and ICM could, on other hand, be well imaged at higher redshifts (beyond $z\sim 0.3$). The results also illustrate the power of a large FoV: even at redshifts as low as 0.03, \textit{HUBS} would still be able to capture a galaxy cluster in its entirety with one pointing, enabling spatially-resolved spectroscopy.

\subsection{Spectral analyses}
\label{sec:fitting}
\begin{figure}[h]
    \centering
    \includegraphics[width=1\textwidth]{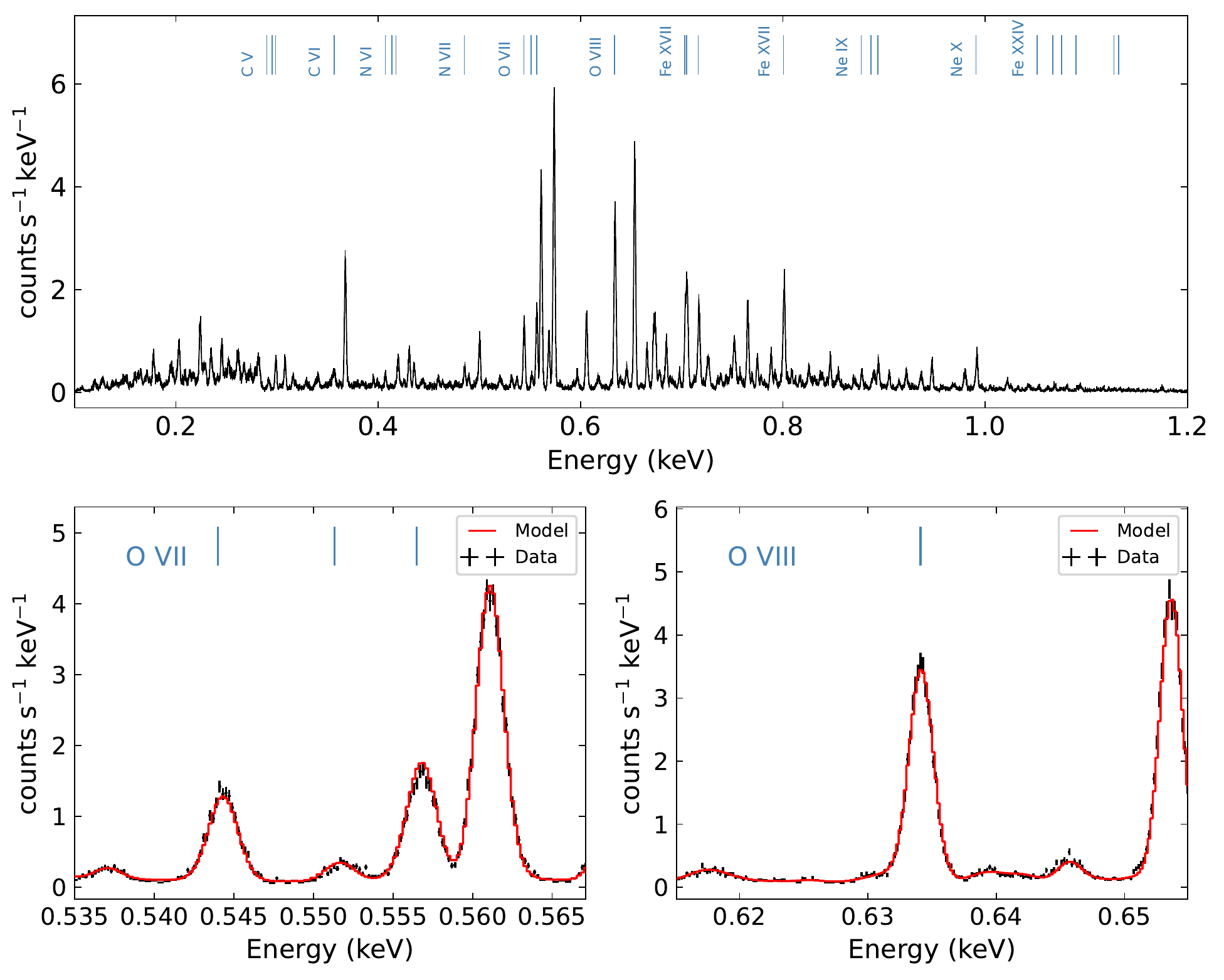}
    \caption{Top panel: Mock X-ray spectrum of Target C. It was made with X-ray photons within a circular region of radius $r_{200}$ centred at the target. At the top, some of the strong emission lines associated with the target are indicated. Bottom panel: Zoomed-in view of the top panel, around the \ion{O}{\romannumeral7} line triplets ({\it left}), and the \ion{O}{\romannumeral8} K$\alpha$ line ({\it right}). Also shown is the best-fitting model (in solid line). The strongest lines are due to the foreground. Some of the weak lines are associated with the light-cone component. } 
    \label{fig:fitting_3}
\end{figure}

To obtain the X-ray spectrum of a target, we collected all photons from within a circular region of radius $r_{200}$ that is centered at the target. Instead of taking a more traditional approach to background subtraction, we chose to include various contributions (the target, light cone, AGN background and Galactic foreground) in the model for spectral fitting. As an example, Fig.~\ref{fig:fitting_3} shows the mock spectrum of Target C. The (redshifted) emission lines expected of the CGM are indicated at the top of the figure. We note that the strong emission lines are associated with the foreground, which could make it challenging for \textit{HUBS} to accurately measure the emission lines associated with an extragalactic source. Fig.~\ref{fig:fitting_3} also shows zoomed-in views of the mock spectrum around the \ion{O}{\romannumeral7} and \ion{O}{\romannumeral8} lines. Besides the emission lines associated with the target, the presence of the foreground (strongest) lines are apparent, along with weak lines likely associated with star-forming galaxies in the light cone. 

Taking the conventionally approach to spectral modeling, we fitted the X-ray spectrum in \textsc{xspec} \cite{1996ASPC..101...17A} with a model that consists of following components: 

\begin{itemize}
    \item Three \verb"bapec" components, approximating the X-ray emission from a multiphase hot gas of the target galaxy;
    \item Two \verb"apec" components, accounting for the Galactic foreground;
    \item One \verb"powerlaw" component, modeling the AGN background (plus other continuum);
    \item A number of \verb"gauss" components, representing the weak emission lines associated with other X-ray sources (e.g. star-forming galaxies) in the light cone.  
\end{itemize}

The interstellar absorption was computed with \verb"tbabs" (also in \textsc{xspec}), assuming a fixed hydrogen column density of $N_{\rm{H}} = 4 \times 10^{20}~\mathrm{cm^{-2}}$. Compared with the \verb"apec" model, the \verb"bapec" model considers the effects of velocity broadening, which is more appropriate for the targets of interest here. In order to make use of the full spectral resolution in the fitting, we adopted \textit{C}-statistics \cite{1979ApJ...228..939C} to provide an unbiased estimate of the model parameters \cite{2009ApJ...693..822H}, allowing low counts ($<10$) in some of the spectral bins. The spectral fitting was carried out over an energy range of $0.1\text{--}2~\mathrm{keV}$. Other than $N_{\rm{H}}$ and the parameters of the foreground, all other parameters in the model were allowed to vary. An example of the best-fitting model is shown in Fig.~\ref{fig:fitting_3}. 

Table~\ref{tab:results_3} shows the key parameters that characterize the properties of the emitting gas (including temperature, metalicity, redshift, and velocity dispersion). Covariance analysis was carried out to assess correlations among the model paramters and to derive confidence intervals of the parameters. We found a significant correlation between the normalization and metalicity in each \verb"bapec" component, but none between other parameter pairs.

\begin{figure}[h]
    \centering
    \includegraphics[width=1\textwidth]{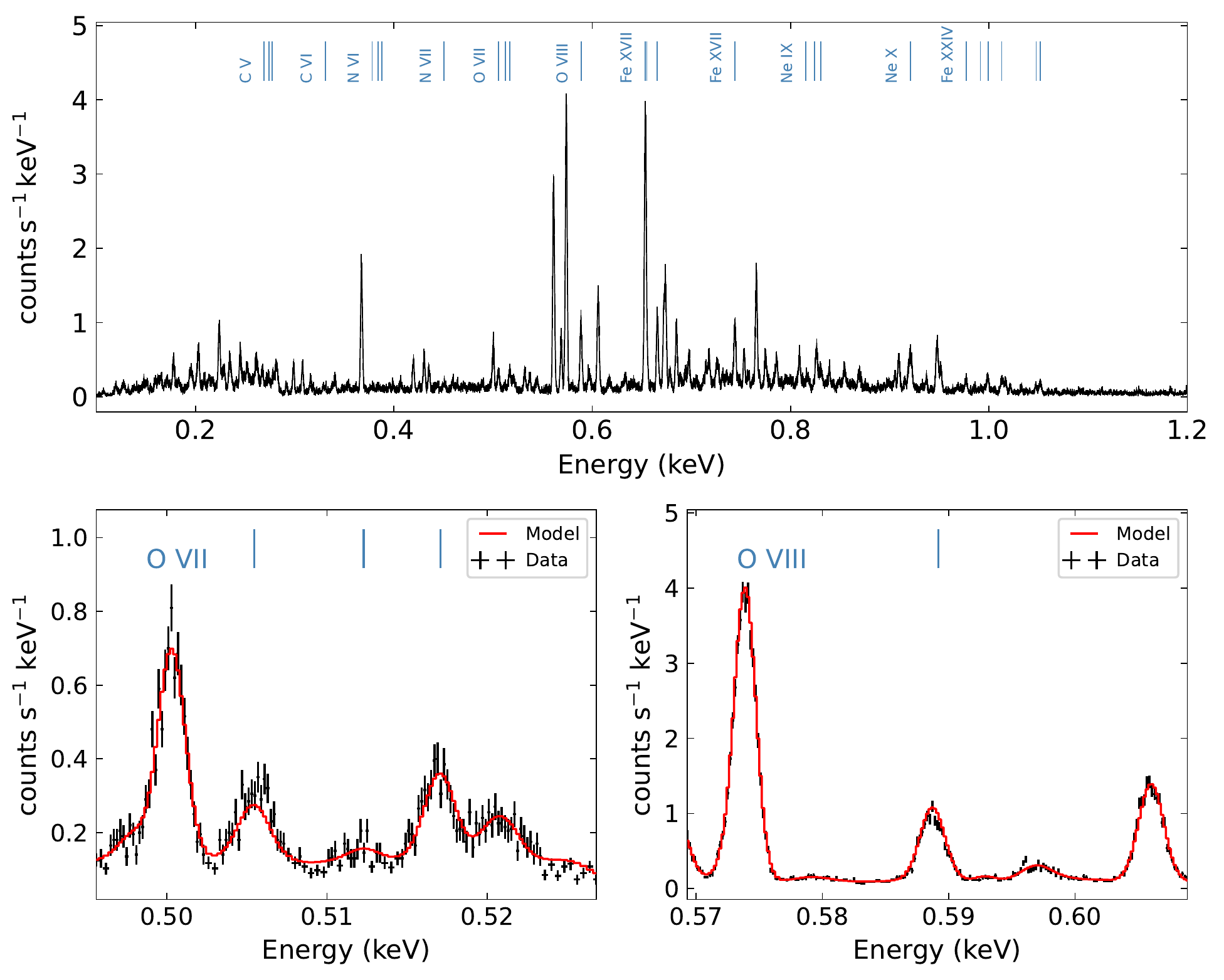}
    \caption{As for Fig.~\ref{fig:fitting_3}, but for Target E.}
    \label{fig:fitting_2}
\end{figure}

\begin{figure}
    \centering
    \includegraphics[width=1\textwidth]{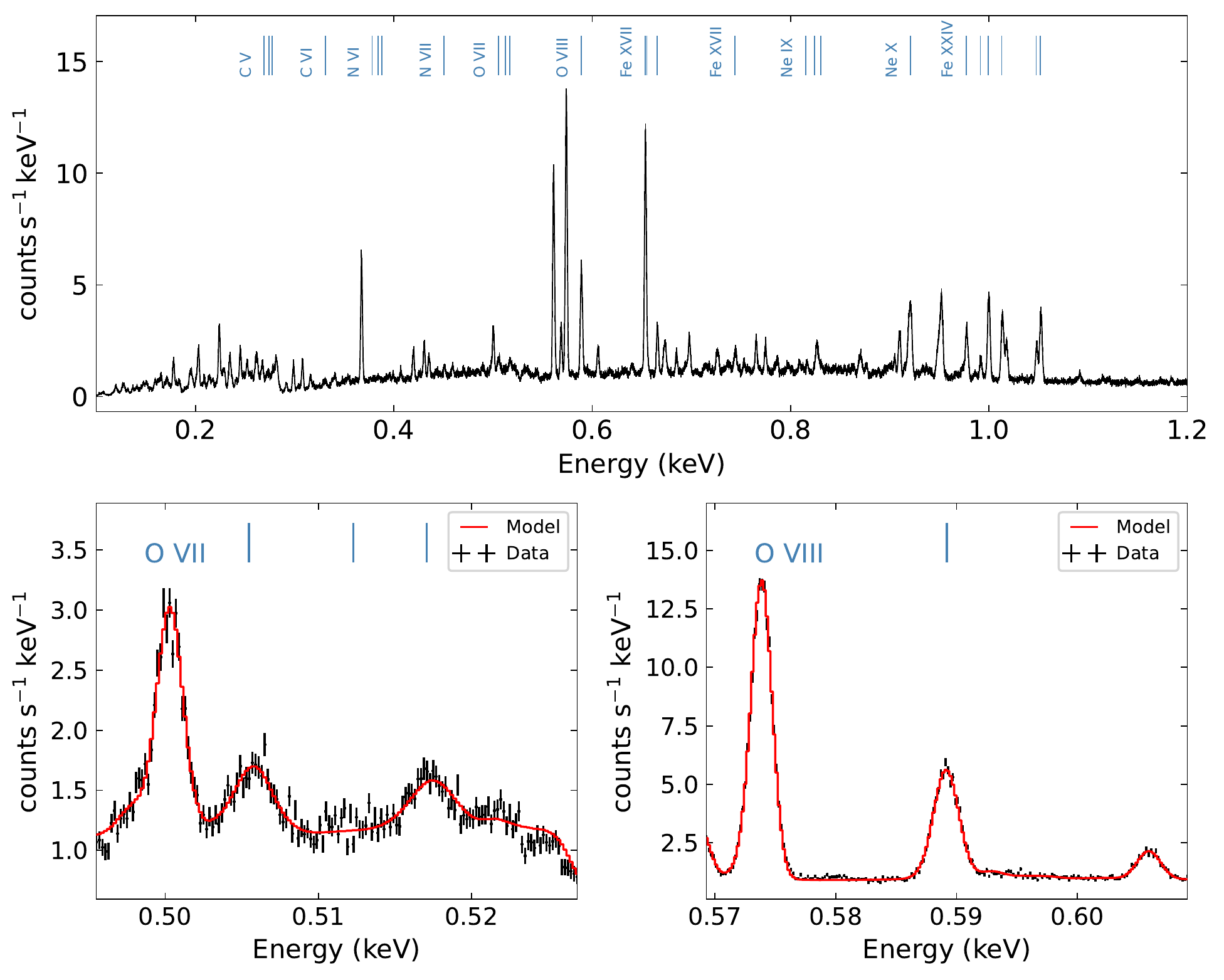}
    \caption{As for Fig.~\ref{fig:fitting_3}, but for Target D. Note the presence of more emission lines associated with the light-cone component, around the \ion{O}{\romannumeral8} line. }
    \label{fig:fitting_1}
\end{figure}

Similar spectral analyses were also carried out for galaxy groups and clusters in the target sample, to measure the properties of IGrM and ICM. The results are also summarized in Table~\ref{tab:results_3}. As examples, Figs.~\ref{fig:fitting_2} and ~\ref{fig:fitting_1} show the overall X-ray spectra of Target E and Target D, respectively.
Zoomed-in views are shown in the figures, focusing on the regions around the \ion{O}{\romannumeral7} and \ion{O}{\romannumeral8} lines. Again, in both cases, deviations from the best-fitting models perhaps indicate possible contribution from contaminating sources in the light cone but, more importantly, inaccuracy in modeling X-ray emission from the hot gas with several discrete \verb"bapec" components (see discussion below). 

\begin{sidewaystable}
\sidewaystablefn%
\begin{center}
\begin{minipage}{\textheight}
	\caption{Results of spectral fitting for the targets}
	\label{tab:results_3}
	\begin{tabular*}{\textheight}{@{\extracolsep{\fill}}cccccc@{\extracolsep{\fill}}} % four columns, alignment for each
      \toprule%
	  Component & Parameter & Unit & Target C & Target E & Target D \\
	  \midrule
      bapec-1 & $kT$ & $\mathrm{keV}$ & $0.426_{-0.005}^{+0.005}$ & $0.712_{-0.007}^{+0.005}$ &  $2.26_{-0.04}^{+0.02}$ \\
      bapec-1 & $Z$ & $\mathrm{Z_{\sun}}$ & $0.59_{-0.03}^{+0.06}$ & $0.066_{-0.002}^{+0.005}$ &  $0.339_{-0.008}^{+0.009}$ \\
      bapec-1 & redshift & & $0.03054_{-0.00001}^{+0.00002}$ & $0.11030_{-0.00005}^{+0.00003}$ &  $0.10944_{-0.00001}^{+0.00001}$ \\
      bapec-1 & $\sigma_\mathrm{v}$ & $\mathrm{km\,s^{-1}}$ & $192_{-10}^{+5}$ & $274_{-14}^{+18}$ & $242_{-3}^{+5}$ \\
      bapec-1 & norm\tnote{a} & $\mathrm{cm^{-5}}$ & $9.8_{-1.0}^{+0.5}\times10^{-5}$ & $6.4_{-0.3}^{+0.1}\times10^{-4}$ &  $5.63_{-0.07}^{+0.13}\times10^{-3}$ \\
      bapec-2 & $kT$ & $\mathrm{keV}$ & $0.197_{-0.001}^{+0.003}$ & $0.250_{-0.008}^{+0.005}$ & $0.659_{-0.010}^{+0.008}$ \\
      bapec-2 & $Z$ & $\mathrm{Z_{\sun}}$ & $0.52_{-0.04}^{+0.04}$ & $0.40_{-0.04}^{+0.05}$ & $0.034_{-0.002}^{+0.002}$ \\
      bapec-2 & redshift & & $0.03078_{-0.00002}^{+0.00002}$ & $0.11003_{-0.00010}^{+0.00014}$ & $0.10955_{-0.00005}^{+0.00008}$ \\
      bapec-2 & $\sigma_\mathrm{v}$ & $\mathrm{km\,s^{-1}}$ & $170_{-14}^{+15}$ & $247_{-27}^{+41}$ & $512_{-17}^{+25}$ \\
      bapec-2 & norm & $\mathrm{cm^{-5}}$ & $1.14_{-0.08}^{+0.10}\times10^{-4}$ & $3.3_{-0.4}^{+0.4}\times10^{-5}$ &  $1.88_{-0.12}^{+0.07}\times10^{-3}$ \\    
      bapec-3 & $kT$ & $\mathrm{keV}$ & $0.871_{-0.008}^{+0.006}$ & $1.31_{-0.01}^{+0.01}$ & $1.205_{-0.018}^{+0.009}$ \\
      bapec-3 & $Z$ & $\mathrm{Z_{\sun}}$ & $0.107_{-0.006}^{+0.009}$ & $0.49_{-0.03}^{+0.06}$ & $2.2_{-0.2}^{+0.2}$ \\
      bapec-3 & redshift & & $0.03066_{-0.00002}^{+0.00005}$ & $0.11047_{-0.00004}^{+0.00004}$ & $0.10933_{-0.00003}^{+0.00003}$ \\
      bapec-3 & $\sigma_\mathrm{v}$ & $\mathrm{km\,s^{-1}}$ & $127_{-14}^{+14}$ & $143_{-12}^{+16}$ & $265_{-11}^{+8}$ \\
      bapec-3 & norm & $\mathrm{cm^{-5}}$ & $2.6_{-0.2}^{+0.2}\times10^{-4}$ & $1.5_{-0.1}^{+0.1}\times10^{-4}$ &  $1.50_{-0.17}^{+0.08}\times10^{-4}$ \\ 
      \botrule
	\end{tabular*}
	\footnotetext{$^a$ $\mathrm{norm}=\frac{10^{-14}}{4\pi \left[D_{\rm{A}} \left(1+z\right)\right]^2} \int n_{\rm{e}} n_{\rm{H}} \, \mathrm{d}V$, where $D_{\rm{A}}$ is the angular diameter distance to the source ($\mathrm{cm}$), $n_{\rm{e}}$ and $n_{\rm{H}}$ are the electron and H densities $\mathrm{\left(cm^{-3}\right)}$, respectively. The uncertainties shown indicate $90\%$ confidence intervals, which were derived from covariance analyses (see the main text). }
\end{minipage}
\end{center}
\end{sidewaystable}

\section{Discussion} \label{sec:summary}

Through analyzing mock observations, it has become fairly clear that, even for a highly-optimized X-ray mission like \textit{HUBS}, it would likely require careful selection of targets and large investment in observing resources to obtain data of decent quality for studying the hot phase of CGM systematically, due primarily to the weakness of the expected signals, and also to the presence of strong foreground emission (and possibly contamination by other sources in the FoV).
What has not been addressed in this work is the problem of separating X-ray emission from the hot CGM and interstellar medium in a target galaxy, owing to the limited angular resolution of \textit{HUBS}. Some of these challenges could be mitigated also through target selection, making use of a wealth of multiwavelength information obtained from prior observations.

Observationally, spectral analysis with \textit{HUBS} data is expected to be quite challenging, due again to the presence of (in some cases) overwhelming foreground lines. Avoiding those lines would also impose constraints on the redshifts. Blending of target lines with foreground lines would complicate reliable modeling. For targets not filling the entire FoV, the standard approach of background subtraction may be effective, although systematic uncertainties could be an issue. At a more fundamental level, it is challenging to model gas with a continuous temperature distribution rigorously in practice; instead, as described in Section \ref{sec:fitting}, it is often approximated with a few discrete components, each of which is characterized by a temperature. The question is how well such an approach can recover the primary phases of the gas. Vijayan \& Li~\cite{2021arXiv210211510V} described modeling with a log-normal distribution in temperature, which seems to provide more reliable results. Novel techniques like this would be required for analyzing X-ray spectra of high resolution in the future.

\subsection{Determining the properties of hot gas}
\label{sec:Determining}

The properties of the hot gas associated with the selected targets have been obtained through spectral analyses. The results can be compared with the properties in the input (simulation).
Specifically, we %tried to compared the original data and the fitting results to 
examine how well the spectral analyses could capture the principal phases of the emitting gas.

\begin{figure}[h]
    \centering
    \includegraphics[width=1\textwidth]{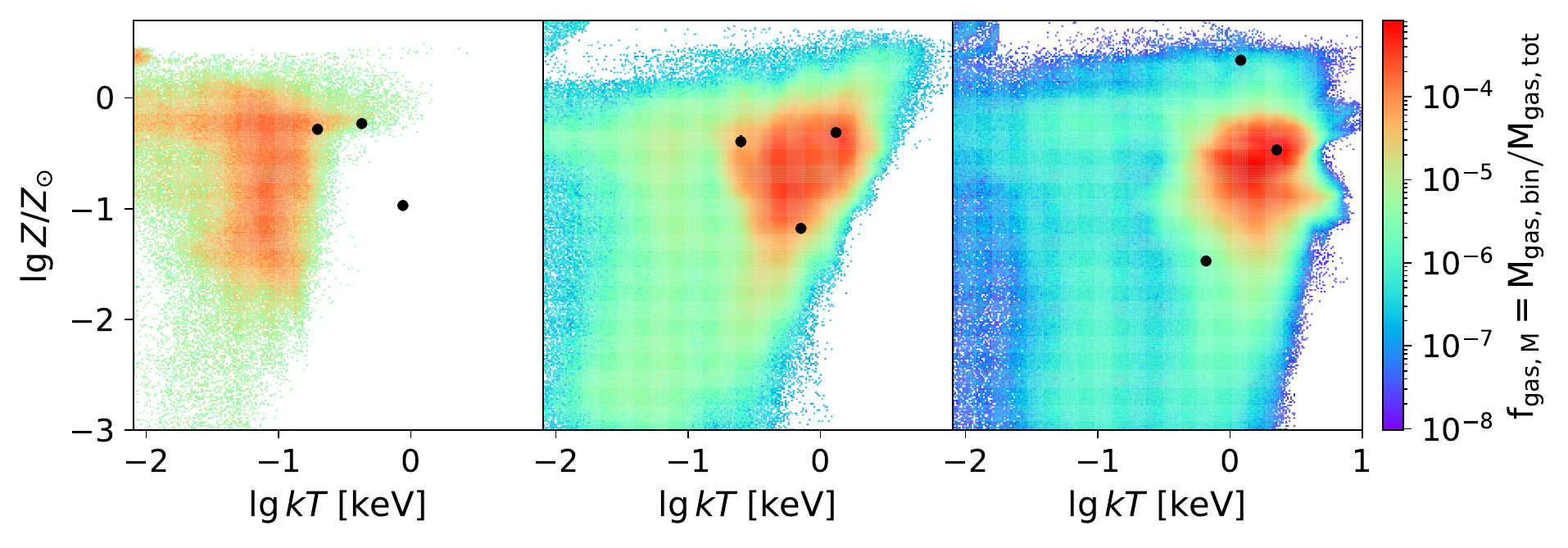}
    \caption{Theoretical phase diagrams ($Z-kT$) for Target C, E and D, respectively. The data were derived from the TNG simulation. For comparison, the results from spectral analyses are shown in black dots. }
    \label{fig:phase}
\end{figure}

Fig.~\ref{fig:phase} shows the theoretical distributions of temperature and abundances (from the TNG data) for Targets C, E and D , respectively.
These were calculated using all gas particles within the chosen spherical region of radius $r_{200}$ for each halo.
Also shown in the figure are the temperature and abundances derived from the spectral modeling. They capture the main components of the multi-phase gas fairly well, but traditional spectral modeling with discrete components clearly shows its limits, partly due to the relatively narrow energy range that \textit{HUBS} is designed to cover. As the figure shows, the approach works very well for the group (Target E), but only marginally for the galaxy (Target C), because the temperature of its CGM is generally quite low and the associated X-ray emission is thus too soft (and weak) even for \textit{HUBS}. At the opposite end, the ICM in the cluster (Target D) seems a bit too hot (thus the associated X-ray emission too hard) for \textit{HUBS} observations to constrain its properties as well as the group.

\subsection{Detection limits}

Of general interest is the sensitivity of \textit{HUBS} to low surface brightness X-ray emission. With the contributions from the light cone, foreground and background components quantified, we can now assess the sensitivity to some of the emission lines of interest (see Fig.~\ref{fig:fitting_3}) fairly realistically. Due to the presence of contaminating lines (associated mainly with the foreground but also with star-forming galaxies along the line of sight), the results derived here are necessarily redshift dependent. We chose $z = 0.03$ for deriving surface brightness limits.

We added X-ray events that are associated with the light cone, Galactic foreground, and AGN background to generate an overall X-ray spectrum, and fitted it with the same model as before (see Section \ref{sec:fitting}). We then added to the best-fitting model a Gaussian component (\verb"zgauss" in \textsc{xspec}), to represent an emission line of interest. We fixed the line energy at the intended (redshifted) value and the line width at $\sigma_K=0.1~\mathrm{eV}$ (accounting for Doppler broadening), and only allowed the normalization of the Gaussian component to vary. We also froze all other model parameters, and fitted the spectrum again, based on \textit{C}-statistics.
The normalization (of \verb"zgauss") was varied until $\Delta C$ provides a $5\sigma$ upper limit. The limiting line surface brightness was then computed. Repeating the procedure, we obtained results for other emission lines of interest. The results are summarized in Table~\ref{tab:sensitivity}.

For comparison, we note that the limiting sensitivity of \textit{Athena} for \ion{O}{\romannumeral7} (r) is about $0.12~\mathrm{photons \, s^{-1} \, cm^{-2} \, sr^{-1}}$ for $10^6~\mathrm{arcmin^2\,s}$ (without considering lightcone contribution) \cite{2021arXiv210804847W}. 

Some caveats should be noted. At low X-ray energies, the Galactic foreground is known to be patchy, so the limiting surface brightness is expected to vary significantly among different lines of sight. In addition, throughout this work, we ignored the detector background due to charged particles in orbit, because it is expected to be negligible.

\begin{table}
	\centering
	\caption{Line surface brightness limits$^\dag$}
	\label{tab:sensitivity}
	\begin{tabular}{lcc}
      \toprule%
      Line & Rest-frame energy & Surface brightness limit \\
      &  $\left(\mathrm{eV}\right)$ & $\left(\mathrm{photons \, s^{-1} \, cm^{-2} \, sr^{-1}}\right)$ \\
      \midrule
      \ion{N}{\romannumeral7} & $500$ & $1.1\times10^{-3}$ \\ 
      \ion{O}{\romannumeral8} & $654$ & $2.1\times10^{-3}$ \\ 
      \ion{O}{\romannumeral7} (r) & $574$ & $3.2\times10^{-3}$ \\ 
      \ion{Fe}{\romannumeral17} & $826$ & $2.4\times10^{-3}$ \\ 
      \ion{Ne}{\romannumeral10} & $1022$ & $1.2\times10^{-3}$ \\ 
      \botrule
	\end{tabular}
	
    $^\dag$ Derived for 5-$\sigma$ detection with an exposure time of 1 Ms.
\end{table}

\section{Summary}
We created mock observations for \textit{HUBS}, based on the IllustrisTNG-100 simulation, and used them to assess the scientific capabilities of \textit{HUBS} in detecting extended X-ray emission from the hot gas in galaxies, groups, and clusters at different redshifts. As expected, the X-ray emission from relatively dense environments (CGM, IGrM, and ICM) can be detected efficiently, at low redshifts, thanks in part to the large FoV of \textit{HUBS}. On the other hand, the emission from filamentary structures in the IGM would seem too faint to be seen (comparing Figs.~\ref{fig:intensity} and \ref{fig:images}). Useful insights into IGM could be gained through studying the outskirt of clusters (perhaps also superclusters). 

For a 2-eV energy resolution of the detector, the emission lines from sources at redshifts lower than about $z=0.01$ become inseparable from those of the foreground emission, although they can still be studied by measuring the foreground and background (and line-of-sight) contribution from the images, as long as they do not fill the entire FoV. The upper redshift bound is determined by the detection sensitivity of \textit{HUBS}, as well as by the brightness of the target. Our results show that it lies around $z\sim 0.2$ for galaxies (see Fig.~\ref{fig:images_oviii}), 
and beyond $z\sim 0.3$ for groups and clusters. Therefore, \textit{HUBS} is mainly for observing targets at relatively low redshifts. It is worth noting that \textit{HUBS} would be well suited for measuring hot IGrM around galaxy groups, and the cooler outskirts of ICM and perhaps also IGM in the cluster environment, with its large FoV. Even at the centre of ICM, \textit{HUBS} would be quite effective in probing the relatively cooler phase of the gas.

The main conclusions of this work are summarised as follows.
\begin{enumerate}
  \item A dedicated X-ray spectroscopic mission similar to \textit{HUBS} would be required to enable significant progress in measuring the spatial distribution and properties of the CGM around nearby galaxies.
  \item Galaxy groups would be excellent targets for \textit{HUBS}, as the hot phase of IGrM peaks around temperatures $kT \sim 1~\mathrm{keV}$.
  \item With its large FoV, \textit{HUBS} would be quite efficient for spectral studies of clusters, especially the outskirts of ICM, even at fairly low redshifts.
  \item At the low end of the accessible redshift range, the Galactic foreground imposes a constraint on the detection of emission lines from an extragalactic source. For a target at a redshift of less than 0.01, its emission lines would be inseparable from the foreground lines, although the traditional background-subtraction method can be applied to the images of sources at even lower redshifts. 
  \item With 1 Ms exposure time, CGM around galaxies could be detected with \textit{HUBS} up to a redshift of about 0.2, IGrM around galaxy groups of to about 0.3, and ICM around galaxy clusters further.
  \item Even with currently available plasma diagnostic tools, \textit{HUBS} data can be used to derive the properties of the primary phases of X-ray-emitting hot gas. More advanced tools would be required to study the multiphase gas more realistically.
\end{enumerate}

In this work, we did not consider absorption-line diagnostics. In the payload design of \textit{HUBS}, though, special consideration has been made to optimize such capability, although the details are still being discussed and may evolve. 

\bmhead{Acknowledgements}

We thank Dan McCammon for providing the \textit{XQC} filter data and for useful discussions, and Zhansan Wang for providing preliminary data on the \textit{HUBS} optics. 
We would also like to thank John ZuHone for useful suggestion on using some of the software packages, and Taotao Fang for advice on making mock observations.
This work was supported in part by the Ministry of Science and Technology of China through its National Key R\&D Program, Grant 2018YFA0404502, and by the National Natural Science Foundation of China through Grant 11821303.
This work made use of several Python packages for astronomy, including \texttt{pyXSIM} \cite{2014arXiv1407.1783Z, 2016ascl.soft08002Z}, \textsc{photutils} \cite{2020zndo...4049061B}, \textsc{astropy} \cite{2013A&A...558A..33A, 2018AJ....156..123A}.
The figures in this paper were made using the python \textsc{matplotlib} \cite{2007CSE.....9...90H} and \textsc{seaborn} \cite{2018zndo...1313201W} package.

\section*{Declarations}

% Some journals require declarations to be submitted in a standardised format. Please check the Instructions for Authors of the journal to which you are submitting to see if you need to complete this section. If yes, your manuscript must contain the following sections under the heading `Declarations':

% \begin{itemize}
% \item Funding
% \item Conflict of interest/Competing interests (check journal-specific guidelines for which heading to use)
% \item Ethics approval 
% \item Consent to participate
% \item Consent for publication
% \item Availability of data and materials
% \item Code availability 
% \item Authors' contributions
% \end{itemize}

% \noindent
% If any of the sections are not relevant to your manuscript, please include the heading and write `Not applicable' for that section. 

% %%===================================================%%
% %% For presentation purpose, we have included        %%
% %% \bigskip command. please ignore this.             %%
% %%===================================================%%
% \bigskip
% \begin{flushleft}%
% Editorial Policies for:

% \bigskip\noindent
% Springer journals and proceedings: \url{https://www.springer.com/gp/editorial-policies}

% \bigskip\noindent
% Nature Portfolio journals: \url{https://www.nature.com/nature-research/editorial-policies}

% \bigskip\noindent
% \textit{Scientific Reports}: \url{https://www.nature.com/srep/journal-policies/editorial-policies}

% \bigskip\noindent
% BMC journals: \url{https://www.biomedcentral.com/getpublished/editorial-policies}
% \end{flushleft}

\subsection*{Funding}

This work was supported in part by the Ministry of Science and Technology of China through its National Key R\&D Program, Grant 2018YFA0404502, and by the National Natural Science Foundation of China through Grant 11821303.

\subsection*{Data Availability}

The IllustrisTNG halo catalogues and the simulation snapshots \cite{2019ComAC...6....2N} are publicly available at \url{http://www.tng-project.org/data/}. The rest of the data underlying the article will be available from the corresponding author on reasonable request.

\bibliography{reference}

%% Default %%
%%\input sn-sample-bib.tex%

\end{document}